\documentclass[twocolumn]{autart}
\usepackage{amsmath}
\usepackage{amssymb}
\usepackage{multirow}
\usepackage{graphicx}          
\usepackage{dirtytalk}
\usepackage{tabularx}
\usepackage{framed} 
\usepackage{multicol} 
\usepackage{etoolbox}
{}
{}
\newtheorem{remark}{Remark}{}
\usepackage{tikz}
\usepackage{pifont}
\usepackage{balance}
\usepackage{multicol, blindtext}

\usepackage{float}
\usepackage[utf8]{inputenc}
\usepackage{lettrine}
\usepackage{eurosym}
\usepackage{adjustbox}
\usepackage{array}
\usepackage{threeparttable}
\usepackage{booktabs}
\usepackage{cuted}
\usepackage{algorithm}
\usepackage{algpseudocode}
\usepackage{url}
\usepackage[none]{hyphenat}
\usepackage{textcomp}

\begin{document}

\begin{frontmatter}

\title{Techno-Economic Modelling and Component Sizing in Renewable Energy Communities: A Participant Perspective}                                       

\author[Benevento]{Vishal Kachhad}\ead{vishalkachhad123@gmail.com},    
\author[Benevento]{Amit Joshi}\ead{aj25021993@gmail.com},              
\author[Naples2]{Luigi Glielmo}\ead{luigi.glielmo@unina.it}        

\address[Benevento]{Department of Engineering, University of Sannio, Benevento, Italy.}
\address[Naples2]{Department of Electrical Engineering and Information Technologies, University of Naples Federico II, Italy.}

\begin{keyword}
sizing; photovoltaic; battery energy storage; energy management; net present value; periodicity                           
\end{keyword}                             

\begin{abstract}                          
This article proposed the optimization problem formulation to find optimal sizes of Photovoltaics (PV) and Battery Energy Storage Systems (BESS) for individual participants within the context of the Renewable Energy Community (REC). An optimization problem considered the dynamic nature of electricity pricing, solar irradiation levels, financial aspects such as capital investment, and operational and maintenance expenditures of PV and BESS. The analysis also considered replacement costs and the efficiency of charging and discharging the BESS unit. We employed Mixed Integer Non-Linear Programming (MINLP) to find the optimal system size to maximize individual participants' Net Present Value (NPV). Furthermore, in this study, we used daily representative signals for each season of the year to reduce simulation runtime. The Fast Iterative Shrinkage-Thresholding Algorithm (FISTA) was used to extract these signals. Then these representative signals obtained were used in the optimization problem formulation to reduce simulation time and extend our analysis to a wider planning horizon. In addition, the study introduced fairness by applying the individual marginal contribution method to distribute incentives equitably among REC participants, ensuring that each member benefited from their contribution. A simulation study was performed using a real demand dataset of five houses located in Roseto Valfortore, a small town and commune of the Foggia Province in the Apulia region of southern Italy, as a proof of the practical relevance and usefulness of the ideas discussed. Ultimately, the goal of the article was to empower the REC with the knowledge necessary to make informed decisions and shape the future of sustainable energy.
\end{abstract}

\end{frontmatter}

\section{Introduction}
\lettrine{T}{he} landscape of renewable energy generation transformed, reflecting the new policies that shaped our power networks. In this era of deregulation, more people and companies became involved in the management of energy resources. An example of this shift was the increasing number of RECs. A REC was a localized group of individuals, businesses, or organizations that collaborated in the generation, distribution, and consumption of renewable energy resources. These communities were characterized by their commitment to sustainable energy practices, with the aim of reducing the reliance on nonrenewable energy sources and minimizing environmental impacts. The key features of an REC included collective ownership or shared access to renewable energy infrastructure, such as solar panels and energy storage technologies. REC participants worked together to exploit natural resources in an environmentally friendly manner, fostering a sense of shared responsibility for energy production and consumption \cite{DEU}. People joined a REC because of the opportunity for real economic advantages, primarily through lower energy bills \cite{JR}. The reduced energy costs and the option to sell extra energy to the grid at reasonable rates added to the savings compared to regular rates. Furthermore, the public authority provided an additional reward for sharing energy within the REC, making it even more appealing to REC members \cite{MK}. This modern concept of the REC was supported by a regulatory framework that allowed participants (at least two) to create fair contracts to share distributed energy resources \cite{SJ}. Within this innovative framework, the REC brought three main benefits. First, it achieved substantial environmental gains, notably in the form of a reduced carbon footprint. Second, it fosters social empowerment by encouraging increased participation from end users, thereby nurturing a sense of community and shared purpose in energy production. Lastly, REC's energy-sharing practices yielded considerable economic benefits, manifested as revenue streams generated through collaborative energy transactions between its members. For these benefits, it was essential to implement intelligent control algorithms and advanced communication technologies that ensured the stable and reliable operation of individual REC units and facilitated potential interactions with external entities, such as aggregators, transmission system operators (TSO), and public authorities.

The control algorithms used in the REC had to align with the rules and regulations provided by the public authority, the preferences of the participants, and the specific operational limits of the REC. Finding the optimal size of the system components in advance was crucial for making the REC profitable, especially when considering how many PV and BESS components were needed. Some earlier studies examined the sizing problems in peer-to-peer energy sharing but did not explicitly consider the revenue earned through energy sharing or the potential savings on transmission costs. For example, a study examined the optimal sizing of solar PV, BESS and diesel generators in \cite{CDR}, while another explored the appropriate sizes for PV, wind and BESS using forecasted power generation data in \cite{RK}. Furthermore, \cite{DA} solved an optimization problem with respect to the sizes of solar PV and BESS installations using data from 300 Australian households. Articles \cite{MA} and \cite{LZ} introduced a model to decide the optimal number of solar PV and BESS units for residential systems, considering uncertainties in solar generation, load demand, and different pricing schemes. Solar PV and BESS served various applications, including peer-to-peer energy trading \cite{PA}, demand side management \cite{XL}, and minimizing electricity bills \cite{YJ}. In \cite{BG}, the authors highlighted various key performance indicators and considered the optimal size of the system from an economic perspective. In \cite{AGC} and \cite{GB}, a two-stage approach and energy management of electrical equipment were explained. However, articles focusing on the REC lacked information on the optimal sizing of PV and BESS for individual participants. In particular, \cite{CGCM} suggested a model for optimal system size to maximize community revenue, and \cite{FDAF} introduced a framework for fair revenue sharing and exit clauses among REC participants, considering the optimal size of solar PV and BESS. The concept of a bilevel model for an internal local market within the REC was explained in \cite{BCIS}, which demonstrated a fair revenue distribution among its members. In addition, some studies considered sizing problems while accounting for the capacity of the distribution network line \cite{WTDD} and the transformer rating \cite{NLFR}. In particular, \cite{APPI} proposed a MINLP scheme for optimal solar PV and BESS sizing, and in \cite{PT}, a bilevel optimization model was used to determine parameters and profiles for new REC participants, considering existing community members. In energy trading, \cite{DMFC} explored energy purchase and bid strategies for energy exchange within and between REC communities.

Although many aspects were well explored, the challenges of optimal system sizing for individual participants needed further investigation to enhance REC’s overall efficiency and effectiveness. Although previous studies addressed sizing issues in RECs, a critical gap existed as they overlooked NPV a crucial factor and confined their investigations to single day profiles, neglecting the longer planning horizon (typically 20 to 25 years). It was imperative to recognize that seasonal variations and uncertainties in energy demand and renewable generation over decades further complicated sizing decisions, necessitating a robust framework that balanced technical feasibility with long-term economic viability. In response to these challenges, this paper presented an innovative approach to optimal sizing of PV and BESS for individual REC participants. Our methodology leveraged representative daily periodic signals of four seasons, spring, summer, fall and winter, blending the dynamic nature of renewable energy sources with the crucial concept of NPV considering the timing of future cash flows. Furthermore, we proposed the individual marginal contribution as a fair and computationally efficient method to allocate costs and benefits proportionally, thus fostering collaboration and stability within the REC. We examined these mechanisms through a numerical study that revealed their characteristics and effectiveness in promoting fairness within the REC. This article aimed to fill critical gaps in existing research by providing information on the optimal sizing of PV and BESS for individual REC participants.

The remainder of the paper was structured as follows. Section II described the mathematical model of the system components; in Section III, we formulate the optimal sizing problem followed by its approximation using the “periodic signal”; Section IV demonstrated the simulation results; and Section V presented the conclusions.

\section{Mathematical Modeling}
\begin{figure}[h!]
    \centering
    \includegraphics[height=6.3cm,width=\linewidth]{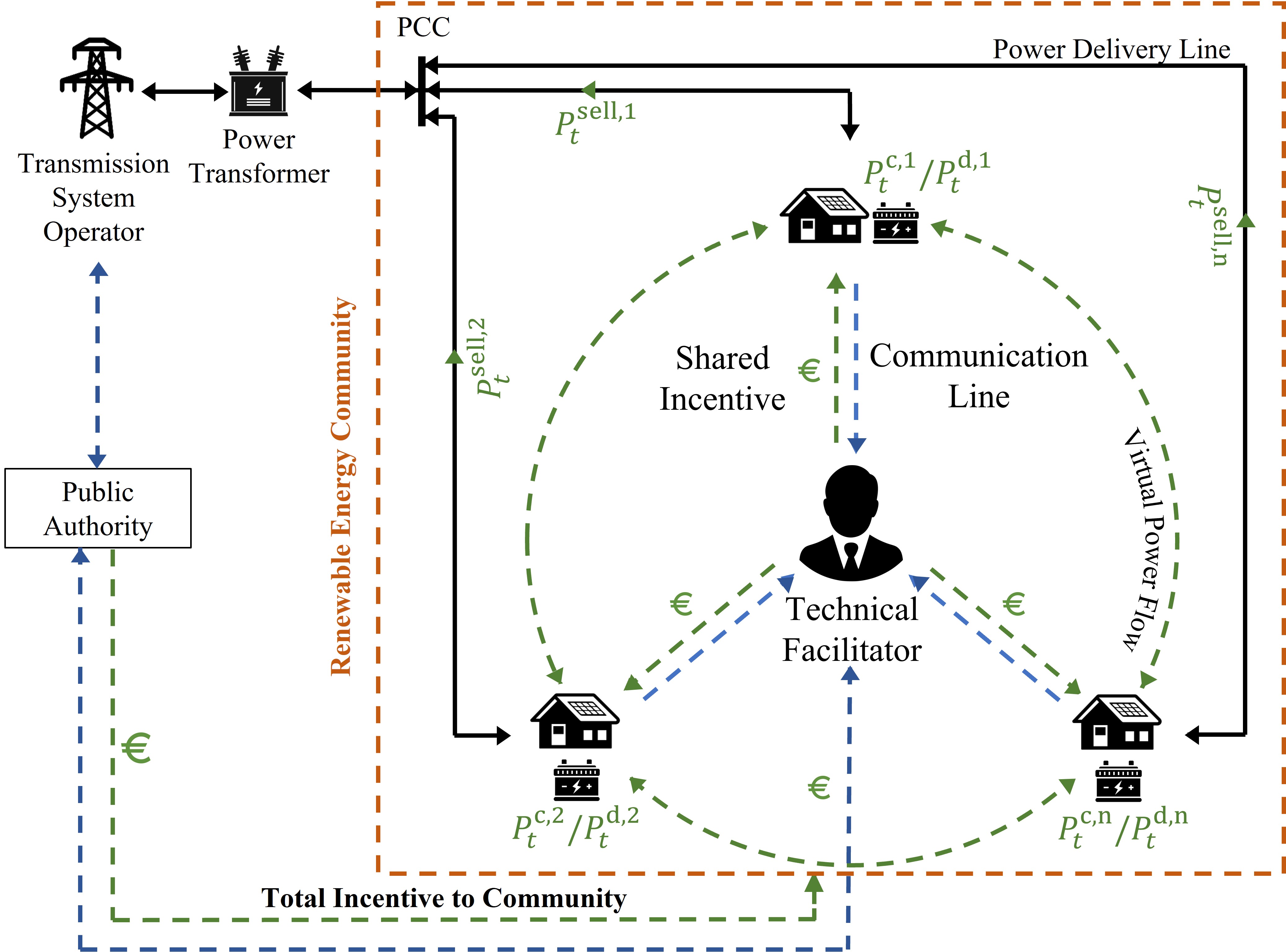}
    \caption{Schematic representation of the REC. In this model, the TF signs a contract with all the participants. This contract covers the real energy management, as well as the operation and maintenance costs for the solar PV and BESS installed on each participant's rooftop.}
    \label{fig:REC_MODEL_II}
\end{figure}

As shown in Fig. \ref{fig:REC_MODEL_II}, we considered a scenario in which $n$ participants and the Technical Facilitator (TF) formed a REC. All participants were equipped with smart meters to facilitate informed decision making by providing real-time information on energy production and consumption. A dedicated TF was an essential entity within the REC to ensure smooth operation and provide technical support to individual participants. The TF played an important role in providing information on the latest technologies, sharing best practices, connecting with resources, and helping navigate technical and regulatory challenges that arose in developing the REC. The TF came from government agencies, nonprofit organizations, or private companies, acting as an essential guide on the journey of building an REC.

The proposed REC model delegated all responsibilities, including design, operation, maintenance, and energy management, to the TF, as mutually decided by all participants. The REC system was monitored and operated under the terms and rewards of the service agreement with the TF. Compensation for TF was specified as a portion of the financial gain obtained by REC participants. On the benefit side, community participants received incentives for sharing energy, reduced electricity bills, and other indirect benefits, such as lower carbon emissions. Following internal regulations and the signed contract, the TF retained a specified amount and the remaining funds were distributed fairly among all participants. Typically, within this framework, the TF made two contracts with the REC participants: (1) a real-time energy management system contract covering the deployment and management of installed solar PV and BESS, and (2) a maintenance and operation costs contract covering the related expenses for each participant based on the number of solar PV and BESS installed.

In addition, each participant maintained a physical connection to the Point of Common Coupling (PCC), which formed the energy exchange network. An individual solar PV captured sunlight and converted it into usable electrical power. This energy played multiple roles within the community: it fulfilled local energy needs, recharged the BESS, sold back to the main grid, and shared within the REC. The transaction was divided into four categories: (1) self-consumed: Each participant leveraged their locally generated solar power to meet their electricity demand and charge the BESS, ensuring efficient self-reliance; (2) energy sold: after self-consumption, any remaining surplus energy was sold to the main grid, contributing to the participants' revenue; (3) energy sharing: all participants met their electricity demand from the solar energy generated within the REC (with a portion of the sold energy constituting the shared energy); and (4) energy bought: participants purchased deficit energy from the main grid during periods of reduced solar generation. In this context, participants were motivated to meet their REC energy demand from locally generated power due to cost savings compared to regular grid prices and incentives received from the public authority.

The detailed discussion and mathematical modeling of different elements of the REC, such as PV energy, BESS, grid, and NPV, are further elaborated. In the following subsections, we describe the mathematical modeling of various REC elements. In this regard, $\mathbb{R}^{+}$, $\mathbb{N}$ and $\lfloor\cdot\rfloor$ denote the set of nonnegative real numbers, the set of natural numbers, and the floor operator, respectively; let $\mathcal{Y}$ be the set of years in the planning horizon and $t\in\mathcal{T}_{y}$ denote a generic time instant corresponding to year $y\in\mathcal{Y}$. Unless otherwise stated, the formulas apply $\forall t\in\mathcal{T}_{y}$ and $\forall y\in\mathcal{Y}$.

\subsection{PV}
PV energy enabled the REC to become a self-sufficient energy generator, reducing the dependence on the main grid. Solar PV offered various benefits beyond its substantial economic advantage, such as playing a crucial role in reducing greenhouse gas emissions and environmental pollution and contributing to a cleaner, more sustainable future. Their presence improved energy security, acting as a valuable power source during power outages. In addition, governments often provided financial incentives or subsidies to improve the affordability of solar PV. For the mathematical modeling, the power generation of a single panel was modeled as per~\cite{YRSF} and was defined as: 
\begin{equation}
    \hat{P}_{t}^{s}=P^{r}\frac{E_{t}}{\hat{E}}\bigg[1 - \frac{\gamma}{100}(\theta_{t}^{c}-\hat{\theta})\bigg], \label{eqn:rated}
\end{equation}
where $E_{t}$, $\gamma$, $P^{r}$ and $\theta_{t}^{c}$ respectively denoted the irradiance level, the power temperature coefficient, the rated capacity of a single panel, and the cell temperature. In addition, $\hat{E}$ and $\hat{\theta}$ were the irradiance and temperature under standard test conditions. Here, the coefficient $\gamma$ was important because it showed how changes in cell or surface temperature affected the power output of solar PVs. In particular, the specific value of $\gamma$ depended on the type of solar PV system employed, whether it was polycrystalline, monocrystalline, thin film, or another variant. This model facilitated a deep understanding of the dynamics that underlay solar power generation and its response to varying environmental conditions. As proposed in~\cite{YRSF}, the cell temperature of solar PV systems ($\theta_{t}^{c}$) was expressed using the following equation:
\begin{equation}
    \theta_{t}^{c}=\theta_{t}^{a}+\frac{E_{t}}{\hat{E}}(\overline\theta - \hat{\theta}),
\end{equation}
where $\overline\theta$ was the Nominal Operating Cell Temperature (NOCT), defined as the temperature reached by open-circuited cells in a module under the conditions: (1) Irradiance on cell surface = 0.8 kW/m$^2$; (2) Air Temperature = 20$^{\circ}$C; and (3) Wind Velocity = 1 m/s. However, it was important to note that since the solar generation determined by \eqref{eqn:rated} applied only to a single PV panel, the $n^{th}$ participant had to install a variable number of PV panels, denoted as $N^{p,n}$, to fulfill their energy requirements. The total solar power generation, taking into account $N^{p,n}$ panels, was defined as:
\begin{equation}
    P_{t}^{s,n} = N^{p,n}\hat{P}_{t}^{s}.
\end{equation}
Moreover, there was also a constraint on the available area for installing solar PV panels. Proper installation demanded a specific amount of space to maximize solar irradiation. This space was often limited in urban environments due to structures like buildings and trees obstructing sunlight. As such, the installation of $N^{p,n}$ solar panels was bounded by the following constraint:
\begin{equation}
    0 \leq N^{p,n} \leq \biggl \lfloor\frac{A^{n}}{A^{p}}\biggr \rfloor. \label{P number}
\end{equation}
In this equation, $A^{n}$ and $A^{p}$ denoted the available space for the $n^{th}$ participant's PV system installation and the area occupied by a single panel, respectively. As the number of solar panels $N^{p,n}$ had to be a whole number, the floor function ($\lfloor \rfloor$) was employed in the computation. In the quest to install these solar panels, each participant procured panels of size $P^{r}$ at a cost of $\hat{C}^{p}$. The total cost for the $n^{th}$ participant's solar panels was calculated as follows:
\begin{align}
    C^{p,n} &= \hat{C}^{p}N^{p,n}. \label{Solar}
\end{align}
Additionally, individual participants within the REC paid the Operating and Maintenance Costs ($\textrm{OPEX}^{p}$) along with the Contract Agreement Costs ($\textrm{CA}^{p}$) associated with solar PV. This combined cost, denoted as $\textrm{OMCA}^{p,n}$, was computed as follows:
\begin{equation}
    \textrm{OMCA}^{p,n} = N^{p,n}\textrm{OPEX}^{p} + N^{p,n}\textrm{CA}^{p}. \label{omtfp}
\end{equation}
For simplicity, the installation assumed that all participants used the same type of PVs.

\subsection{BESS}
The BESS was a valuable asset for all REC participants, allowing them to save substantially on their electricity bills. This was made possible by optimally capturing locally generated energy during high-demand periods with high electricity rates. Later, this stored energy was used to meet demand when it was most cost-effective. In addition, the BESS served as a reliable backup power source during unexpected outages, adding additional resilience to the REC. The state of charge of a single BESS was modeled using the energy-reservoir model~\cite{CDR}, which was characterized by charging and discharging efficiencies:
\begin{equation}
    \hat{S}_{t+ \Delta t} = \hat{S}_{t}+\bigg[\eta^{c}\hat{P}_{t}^{c}-\frac{\hat{P}_{t}^{d}}{\eta^{d}}\bigg]\Delta t, \label{SOC1}
\end{equation}
where $0 \leq \hat{P}_{t}^{c}\leq \overline{P}^{c}$ and $0\leq \hat{P}_{t}^{d}\leq \overline{P}^{d}$ were the bounded charging and discharging power, and $\eta^{c}$ and $\eta^{d}$ respectively denoted the charging and discharging efficiencies of the BESS. As the capacity of the BESS was finite, the state of charge was also bounded, i.e.,
\begin{equation}
\underline{S} \leq \hat{S}_{t} \leq \overline{S}, \label{eqn:SOC1}
\end{equation}
where $\underline{S}$ and $\overline{S}$ were the minimum and maximum states of charge of the BESS. Let $P_{t}^{c,n}$ and $P_{t}^{d,n}$ denote the aggregate charging and discharging power for the $n^{\rm{th}}$ participant associated with the number of BESS installed, i.e., $N^{b,n}$, such that
\begin{subequations}
\label{eqn:ChargeDischarge_REC2}
\begin{align}
0 \leq P_{t}^{c,n} &\leq N^{b,n}\overline{P}^{c}b_{t}^{c,n}, \label{Charge}\\[1mm]
0 \leq P_{t}^{d,n} &\leq N^{b,n}\overline{P}^{d}b_{t}^{d,n}, \label{Discharge}
\end{align}
\end{subequations}
where $b_{t}^{c,n}$ and $b_{t}^{d,n}$ were binary variables imposed to satisfy the non-simultaneous charging and discharging of the BESS by enforcing the constraint
\begin{equation}
    b_{t}^{c,n} + b_{t}^{d,n} \leq 1.  \label{binary}
\end{equation}
Accordingly, \eqref{eqn:SOC1} was modified as 
\begin{equation}
N^{b,n}\underline{S} \leq S_{t}^{n} \leq N^{b,n}\overline{S}, \label{eqn:SOC2}
\end{equation}
where 
\begin{equation}
    S_{t+ \Delta t}^{n} = S_{t}^{n} + \bigg[\eta^{c}P_{t}^{c,n}-\frac{P_{t}^{d,n}}{\eta^{d}}\bigg]\Delta t. \label{eqn:SOC3}
\end{equation}
Let us introduce $\hat{C}^{b}$ as the cost of a BESS with a capacity of $B$. Consequently, the total expenditure for procuring $N^{b,n}$ BESS units for each $n^{\rm{th}}$ participant was calculated as per the following equation:
\begin{equation}
    C^{b,n}=\hat{C}^{b}N^{b,n}. \label{Battery}
\end{equation}
The decision variable of the BESS units, represented by $N^{b,n}$, also had to adhere to an upper limit, denoted $\overline{N}^{b}$, to suit the specific requirements of the systems. Consequently, the imposed constraint was defined as follows:
\begin{gather}
0 \leq N^{b,n} \leq \overline{N}^{b}.  \label{B number}
\end{gather}
Additionally, individual participants within the REC were responsible for covering Operating and Maintenance Costs ($\textrm{OPEX}^{b}$) along with Contract Agreement Costs ($\textrm{CA}^{b}$) associated with their BESS units. This collective cost, denoted as $\textrm{OMCA}^{b,n}$, was computed as follows:
\begin{equation}
    \textrm{OMCA}^{b,n} = N^{b,n}\textrm{OPEX}^{b} + N^{b,n}\textrm{CA}^{b}. \label{omtfb}
\end{equation}
To streamline the analysis, it was assumed that all participants in REC used the same type of BESS.

\subsection{Power Flows} 
In our considered REC setup illustrated in Figure~\ref{fig:REC_MODEL_II}, $P_{t}^{sell,n}$ denoted the surplus power sold to the grid by the $n^{th}$ participant at time $t$, $P_{t}^{sh,n}$ was the power shared by the $n^{th}$ participant with other members of the REC at time $t$ (treated as a virtual quantity), $P_{t}^{c,n}$ and $P_{t}^{d,n}$ represented the charging and discharging power of the BESS belonging to participant $n$ at time $t$, and $P_{t}^{self,n}$ was the power that participant $n$ consumed from their own renewable source at time $t$. The power balance constraints, as defined in \eqref{eqn:Loadbalance_REC2}, ensured that the power supply consistently met the demands of the participants at all times.
\begin{subequations}
\label{eqn:Loadbalance_REC2}
\begin{align}
&P_{t}^{sell,n} = P_{t}^{s,n} - P_{t}^{self,n}, \label{eqn:Load1}\\
&P_{t}^{self,n} \leq P_{t}^{s,n}, \label{eqn:Load2}\\
&P_{t}^{self,n} \leq d_{t}^{n} + P_{t}^{c,n} - P_{t}^{d,n}, \label{eqn:Load3}\\
&P_{t}^{\textrm{sh}} \leq \sum_{n\in \mathcal{N}} \Bigl(d_{t}^{n} + P_{t}^{c,n} - P_{t}^{d,n} - P_{t}^{self,n}\Bigr), \label{eqn:Load4}\\
&P_{t}^{\textrm{sh}} \leq \sum_{n\in \mathcal{N}} P_{t}^{sell,n}, \label{eqn:Load5}\\
&P_{t}^{c,n} \leq P_{t}^{s,n}, \label{eqn:Load6}
\end{align}
\end{subequations}
where $P_{t}^{s,n}$ and $d_{t}^{n}$ respectively denoted the total solar power generation at time $t$ by the $n^{th}$ participant from $N^{p,n} \in \mathbb{N}$ number of panels and the load demand of the $n^{th}$ participant at time $t$. Furthermore, $P_{t}^{self,n}$ denoted the power that the $n^{th}$ participant consumed from their own renewable resources. All these power flows had to be positive to ensure that our optimization model was realistic. Therefore, we imposed the following constraint:
\begin{equation}
    P_{t}^{sell,n},~~ P_{t}^{sh},~~ P_{t}^{self,n}  \geq 0. \label{Positive}
\end{equation}
This constraint guaranteed that there were no negative power values, which helped maintain a clear and feasible energy management model.

\subsection{Electricity Rate}
An electricity rate, a pivotal factor in the sector of energy economics, played a dual role in determining the financial dynamics of participants within the electricity market. It was both the cost borne by participants for their consumption and the revenue they earned when selling surplus electricity back to the grid. The electricity providers decided on this rate, which was subject to various variables that contributed to its fluctuation. Factors such as time of day, the prevailing season, and overall electricity demand influenced the changes in electricity rates. Typically, prices peaked during the summer, when the overall demand for electricity increased. This occurred because of the necessity of incorporating more expensive generation sources to cope with the increased energy requirements during that period. It was essential to recognize that the participants' purchase price was higher than the selling price, reflecting the higher distribution costs incurred in delivering electricity to end consumers \cite{ARERA}.

\begin{figure}[h!]
    \centering
    \includegraphics[width=\linewidth]{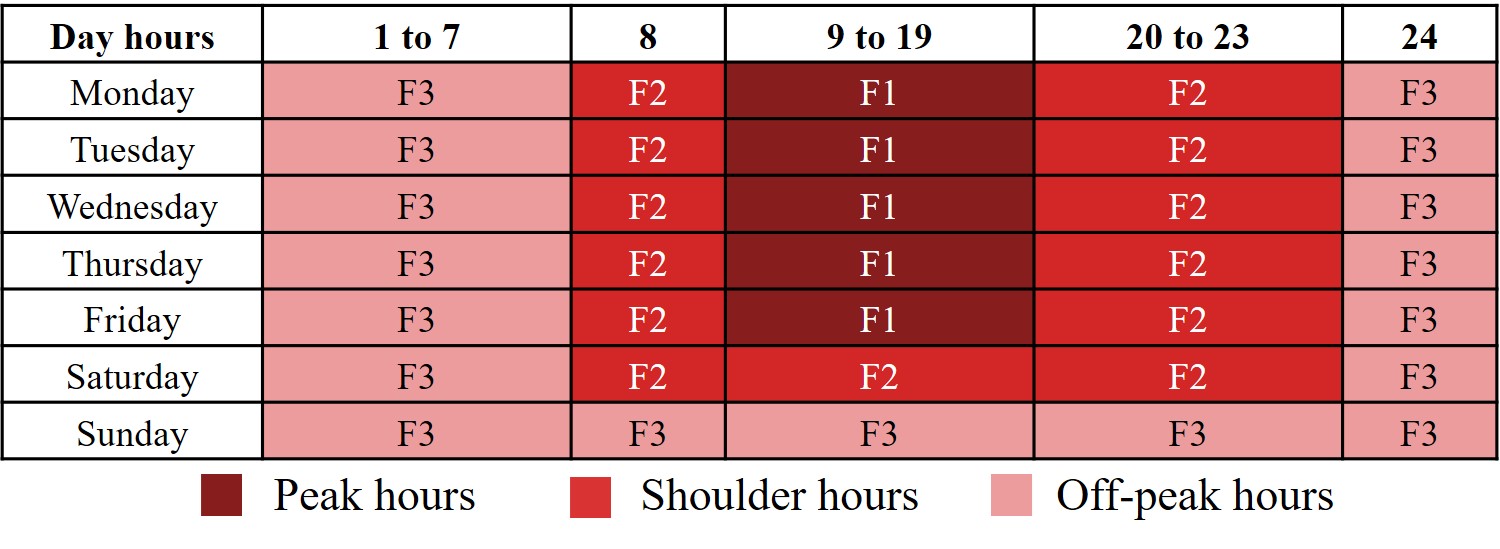}
    \caption{Italian grid electricity rate structure.} 
    \label{fig:ELECTRICITY_RATES}
\end{figure}

As shown in Figure \ref{fig:ELECTRICITY_RATES}, and focusing on the specific context of Italy, the daily electricity price was divided into three distinct bands. F1, F2, and F3. Here, F1 covered peak hours from Monday to Friday between 9 and 19, with the highest rates due to increased demand and the use of more expensive generation sources. F2 included shoulder hours, covering the morning and evening periods on weekdays, 8 and 20 to 23 and Saturdays, 8 to 23, with rates lower than F1 but slightly higher than F3. Lastly, F3 represented off-peak hours, primarily during the night on weekdays from 24 to 7 and throughout the day on Sundays from 1 to 24, with the lowest electricity rates. Additional details on different costs, revenue, and incentives for the considered REC systems were further explained.

Let us consider that the power $P_{t}^{sell,n}$ was sold to the grid by the $n^{th}$ participant at a price $C_{t}^{S}\in\mathbb{R}^{+}$; then the total revenue for power selling to the grid for the $n^{th}$ participant was denoted as
\begin{equation}
    R^{sell,n} = \sum_{t\in \mathcal{T}} C_{t}^{S}P_{t}^{sell,n}  \Delta t.   \label{Revenue}
\end{equation}
As collective solar generation $P_{t}^{s,n}$ served the dual role of meeting the local load demand $d_{t}^{n}$ and supplying power $P_{t}^{c,n}$ for the charging of the BESS, it entailed financial benefits for individual participants. This dual functionality translated into savings as they offset their electricity costs at the prevailing price $C_{t}^{P}$. Consequently, the total savings attributed to self-consumed power for the $n^{th}$ participant was expressed as follows:
\begin{equation}
    R^{self,n} = \sum_{t\in \mathcal{T}} C_{t}^{P}P_{t}^{self,n} \Delta t.   \label{Savings}
\end{equation}
In instances where a participant found themselves with surplus solar generation, they engaged in energy sharing, thereby establishing a mutually beneficial transaction with another REC participant. This exchange occurred at a specified rate, denoted as $C_{t}^{E}$. It should be noted that this rate of sharing energy, $C_{t}^{E}$, was greater than the rate at which excess energy was sold to the grid, $C_{t}^{S}$. This distinctive pricing structure yielded an additional revenue stream for the REC participant and actively encouraged the use of locally generated solar power. Furthermore, when energy was consumed within the REC, it conferred the advantage of economizing on transmission power costs, marked by the rate $C_{t}^{T}$, another benefit provided by the public authority. This dual advantage further contributed to the growing popularity of REC development. The combined incentive for the $n^{th}$ participant was summarized as follows:
\begin{equation}
    I^{sh} = \sum_{t\in \mathcal{T}}(C_{t}^{E} + C_{t}^{T})P_{t}^{\textrm{sh}}  \Delta t.   \label{Incentive}
\end{equation}

\subsection{NPV}
The NPV was the sum of the present value of future cash flows (CF) for the accounting period associated with the investment, and it was discounted so that it was expressed in today’s worth of money \cite{MMM}. In other words, the money earned in the future was considered less valuable at the initial time of investment. That is, by denoting the cash flows at year $y$ as CF$_{y}$ and the discount rate as $r$, the generic NPV for any system was written as follows:
\begin{equation}
    \textrm{NPV} = -\textrm{INV} + \sum_{y \in \mathcal{Y}} \frac{\textrm{CF}_{y}}{(1 + r)^{y}},  \label{NPV}
\end{equation}
where $\textrm{INV}$ was the total investment cost. In the presented modeling, NPV was calculated for the individual participants of the REC. For the calculation, the incentive on sharing power within the REC $I^{sh}$, revenue for selling power to the grid $R^{sell,n}$, and savings on self-consumption $R^{sav,n}$ were used to estimate the future CF of each year $y$ associated with the initial time of investment. Then, the cumulative computation of NPV was performed by adding up each year’s CF over the entire planning horizon. From \eqref{omtfp}, \eqref{omtfb}, \eqref{Revenue} and \eqref{NPV}, the total NPV for the $m^{th}$ year was estimated using the formula:
\scriptsize	
\begin{equation}
    \textrm{NPV}_{m}^{n} = \sum_{y = 0}^{m}\frac{\begin{aligned}-\textrm{INV}_{y}^{n} + R_{y}^{sell,n} + R_{y}^{self,n} + \zeta^{n} I_{y}^{sh} \\ - \textrm{OMCA}^{p,n} - \textrm{OMCA}^{b,n}\end{aligned}}{(1 + r)^{y}}, \label{Actual}
\end{equation}
\normalsize
where $\zeta^{n}$ was the distribution factor of individual participants, which was obtained using the formula derived in \eqref{eqn:Distribution_Factor}. Also, $\textrm{INV}_{y}^{n}$ denoted the investment by the $n^{th}$ participant corresponding to the $y^{\textrm{th}}$ year, defined as
\begin{equation}
    \textrm{INV}_{y}^{n} = 
    \begin{cases}
    C^{p,n} + C^{b,n}, & y = 0 \quad (\textrm{Initial investment}),\\[2mm]
    C^{b,n}, & y = L^{b} \quad (\textrm{BESS was replaced}),\\[2mm]
    0, & \textrm{Otherwise.}
    \end{cases} \label{INVV}
\end{equation}
The payback period for the $n^{th}$ participant was calculated based on \eqref{Actual}, which provided information about the duration of the time period in which the investment money was recovered by selling energy to the grid, receiving incentives on energy sharing, and saving on self-consumption. It was written using the following equation:
\begin{equation}
    \textrm{PB}^{n} = \min\{ m ~|~ \textrm{NPV}_{m}^{n} \geq 0, \, \forall m \in [m, |\mathcal{Y}|] \},
    \label{eqn:payback}
\end{equation}
where the cardinality of $\mathcal{Y}$, denoted as $\lvert \mathcal{Y} \rvert$, was the length of the planning horizon, which included all the periods considered for the investment evaluation. It was often used as a simple way to evaluate the profitability and risk of an investment. After the payback period, the investment continued to generate cash flow for the remainder of the lifespan of the PV-BESS systems, which was considered income generated by the investment. Finally, the net profit (NP) of the entire REC was calculated based on the following equation:
\begin{equation}
    \textrm{NP} = \sum_{n \in \mathcal{N}}\textrm{NPV}_{\lvert \mathcal{Y} \rvert}^{n}, \label{netprofit1}
\end{equation}
where the NP comprised all future cash flows discounted back to their present value, thus providing a comprehensive measure of the financial performance of the REC investment over its lifespan.

\section{Optimal Sizing Problem Formulation}
For the considered REC systems, an optimization problem was formulated to find the optimal number of solar PV and BESS for individual participants in the REC, which involved the maximization of NP while meeting a set of constraints. The constraints included factors such as total power exchange with the grid, the maximum power output of the PV panels and BESS, the capacity of the BESS, and other design constraints.

For the considered REC system with $N$ participants, the optimization problem was formulated as follows:
\begin{equation}
\begin{aligned}
    & \max_{\mathcal{X}}
    && \underbrace{\textrm{\eqref{netprofit1}}}_\text{Net Profit of REC}\\
    & \textrm{s.t.}
    && \underbrace{\textrm{\eqref{eqn:Loadbalance_REC2}}}_\text{Power balance constraints}\\
    &&& \underbrace{\textrm{\eqref{P number}, \eqref{eqn:ChargeDischarge_REC2} to \eqref{eqn:SOC3}, \eqref{B number}, \eqref{Positive}, \eqref{eqn:payback}}}_\text{Operational constraints}\\
\end{aligned}
\label{eqn:OP_1}
\end{equation}
where $\mathcal{X} = \{N^{p,n}, N^{b,n}, P_{t,y}^{self,n}, P_{t,y}^{sell,n}, P_{t,y}^{sh,n}, P_{t,y}^{c,n}, P_{t,y}^{d,n}, \\ S_{t,y}^{n} ~|~ \forall t \in \mathcal{T}_{y}, \forall y \in \mathcal{Y}, \forall n \in \mathcal{N}\}$ was the set of decision variables and $\mathcal{T}_{y}:= \{1,2,\dots, T_{y}\}$, $\forall y \in \mathcal{Y}$ was the planning horizon. The yearly horizon was typically the same for all years, that is, $T_y = \bar{T}, \ \forall y \in \mathcal{Y}$.

\begin{remark}
We aim to maximize NP by finding the optimal value of each variable that produced the highest output value. Solving \eqref{eqn:OP_1} for a planning horizon of the order of decades was computationally very heavy due to the increase in the number of decision variables and constraints that had to be satisfied at each time instant. As an alternative to this, we proposed to take advantage of the periodicity of solar generation and the demand profile, as described in the following subsection.
\end{remark}

\subsection{Exploiting Periodicity}
Solar power generation and demand profiles in a house had inherent periodicities that were used to reduce the dimensionality of the optimal sizing problem and the execution runtime. Although the periodicity could have been daily, weekly, monthly, or yearly, for simplicity the daily periodic signal was used for the analysis since it covered all working hours of the day. In parallel to leveraging daily periodicity, the electricity demand profiles for REC participants were preprocessed using logarithmic transformation and detrending to capture their multi-seasonal characteristics \cite{SR}, \cite{JRR}. More precisely, the following model for the load (electricity demand) $L(t)$ was assumed:
\begin{equation}
    L(t) = \exp\bigg(T(t) + F(t) + \sum_{i}H_{i}(t) + e(t)\bigg)
\end{equation}
where $T(t)$ denoted the trend, $F(t)$ was the deterministic multi-seasonal component including daily, weekly, and yearly periodicities, $H_{i}(t)$ were the intervention events accounting for the effect of holidays, and $e(t)$ was an error term. Following \cite{AG}, the trend $T(t)$ was estimated as the output of a lowpass filter with bandwidth $\frac{1}{730}$ day$^{-1}$. The log-transformed and detrended series
\begin{equation}
    y(t) = \ln(L(t)) - T(t) = F(t) + \sum_{i}H_{i}(t) + e(t)
\end{equation}
exhibited a relatively stable profile, whose variability was mostly due to the simultaneous presence of daily, weekly, and yearly seasonalities. To calculate the multi-seasonal component $F(t)$, a Fourier expansion for weekly and yearly periodic signals was used as
\begin{subequations}
\label{eqn:Fourier}
\begin{align}
    F_{y}(t) &= a_{0} + \sum_{j=1}^{N_{y}}a_{j}\cos{(j\Psi t)} + b_{j}\sin{(j\Psi t)}, \quad \Psi = \frac{2\pi}{T_{y}},\\[2mm]
    F_{w}(t) &= \hat{a}_{0} + \sum_{k=1}^{N_{w}}\hat{a}_{k}\cos{(k\Omega t)} + \hat{b}_{k}\sin{(k\Omega t)}, \quad \Omega = \frac{2\pi}{T_{w}}.
\end{align}
\end{subequations}
An interaction between weekly and yearly periodic signals was taken into account for the calculation of the final multi-seasonal component $F(t)$ using the equation
\begin{equation}
    F(t) = \sum_{i=1}^{2N_{w}(1+2N_{y})}\theta_{i}h_{i}(t), \quad h_{i}(t) \in \mathcal{Y} \otimes \mathcal{W},
\end{equation}
where $\mathcal{Y} = \{\cos(j\Psi t),\, j \in [0,N_{y}]\} \cup \{\sin(j\Psi t),\, j \in [1,N_{y}]\}$ and $\mathcal{W} = \{\cos(k\Omega t),\, k \in [0,N_{w}]\} \cup \{\sin(k\Omega t),\, \\ k \in [1,N_{w}-1]\}$. Here, $N_{y}$ and $N_{w}$ were the numbers of yearly and weekly harmonics, respectively. These selected numbers of harmonics allowed the model to capture variations associated with different times of the week, reflecting weekly cycles within the larger yearly cycle. Let $\theta \in \mathbb{R}^{n}$ denote the vector that contained the coefficients of the Fourier expansion of $F(t)$. Then,
\begin{equation}
    \textrm{Y} = \Phi\theta + \varepsilon, \label{Linear}
\end{equation}
where $Y = [y(1)\, y(2) \dots y(n)]^{T}$ was the vector of log-transformed and detrended loads, and the columns $\phi_{i}$ of $\Phi \in \mathbb{R}^{n\times m}$ were the sampled regressors, i.e., $\phi_{i} = [h_{i}(1) \, h_{i}(2) \dots h_{i}(n)]^{T}$. For the model in \eqref{Linear} with $Y \in \mathbb{R}^{n\times 1}$, $\Phi \in \mathbb{R}^{n \times m}$, $\theta \in \mathbb{R}^{m\times 1}$ and $\varepsilon$ as a noise term, the LASSO regularized estimate \cite{LASSO} was defined as:
\begin{equation}
    \theta^{\textrm{LASSO}} = \underset{\theta}{\arg\min}\bigg(\frac{1}{n}\lVert \Phi\theta - \textrm{Y}\rVert^{2}  + \lambda \lVert \theta \rVert_{1}\bigg), \label{eqn:LASSO}
\end{equation}
where the balance between the loss function and the regularization term was controlled by the scalar regularization parameter $\lambda$. To solve \eqref{eqn:LASSO}, the FISTA method \cite{FISTA} was used, as described in Algorithm~\ref{alg:two}.

\begin{algorithm}[H]
\caption{FISTA Algorithm}\label{alg:two}
\begin{algorithmic}[1]
\State Initialize $\theta_{0}$
\State Set $t_{0} = 1$
\State $\gamma \gets \frac{1}{\textrm{L}_{f}}$ \Comment{$L_{f}$ denotes the Lipschitz constant}
\While{$\lVert \theta_{k} - \theta_{k-1} \rVert > \textrm{tol}$}
    \State $\nabla E(\theta_{k-1}) \gets 2\Phi^{T}(\Phi\theta_{k-1} - Y)$ \Comment{Gradient term}
    \State $\theta_{k} \gets \theta_{k-1} - \gamma \nabla E(\theta_{k-1})$
    \State $t_{k} \gets \frac{1+\sqrt{1+4t_{k-1}^{2}}}{2}$ \Comment{Momentum term}
    \State $\theta_{k} \gets \theta_{k-1} + \frac{t_{k-1} - 1}{t_{k}}(\theta_{k-1} - \theta_{k-2})$ \Comment{Nesterov's acceleration}
    \State $\theta_{k} \gets \textrm{Prox}_{\gamma \lambda}(\theta_{k})$ \Comment{Proximal operator}
\EndWhile
\end{algorithmic}
\end{algorithm}

Here, $\gamma$ was the Lipschitz constant, $\nabla E$ denoted the gradient, $t_{k}$ was the momentum term, $\theta_{k}$ was the updated parameter vector using Nesterov's acceleration to improve convergence speed, and $\textrm{Prox}_{\gamma \lambda}$ was the proximal operator defined as
\begin{equation}
    \textrm{Prox}_{\gamma \lambda}(\theta_{k}) =
    \begin{cases}
    \theta_{k} - \gamma \lambda, & \textrm{if} \ \theta_{k} > \gamma \lambda, \\
    0, & \textrm{if} \ \theta_{k} \in [-\gamma \lambda, \gamma \lambda], \\
    \theta_{k} + \gamma \lambda, & \textrm{if} \ \theta_{k} < -\gamma \lambda.
    \end{cases}
\end{equation}
In our study, a yearly pattern of electricity usage was analyzed to explore its daily trends. A typical signal spanned one year was extracted using the FISTA method and a single day from each season of the year was selected. From these four representative daily signals, denoted as $x$, the NPV was recalculated as
\scriptsize	
\begin{equation}
    \overline{\textrm{NPV}}_{m}^{n}  = \sum_{y = 0}^{m}\frac{\begin{aligned}-\textrm{INV}_{y}^{n} + \beta \bigl(R_{x}^{sell,n} + R_{x}^{self,n} + \zeta^{n} I_{x}^{sh}\bigr) \\ - \textrm{OMCA}^{p,n} - \textrm{OMCA}^{b,n}\end{aligned}}{(1 + r)^{y}}, \label{forpoc1}
\end{equation}
\normalsize
where $x$ represented the obtained periodic signals for the four seasonal days; $R_{x}^{sell,n}$, $R_{x}^{sav,n}$, and $I_{x}^{sh}$ respectively denoted the revenue for selling power to the grid, the savings on self-consumption, and the incentive on energy sharing within the REC for the representative periodic seasonal days of the $n^{th}$ participant. To obtain approximate annual income and costs, the sum was multiplied by $\beta = 91$, which was the total number of days in a season. From \eqref{forpoc1}, the NP of the entire REC was calculated using the expression
\begin{equation}
    \overline{\textrm{NP}}  = \sum_{n \in \mathcal{N}}\overline{\textrm{NPV}}_{\lvert \mathcal{Y} \rvert}^{n}. \label{forpoc2}
\end{equation}  

\subsection{Optimal Sizing using Periodic Signals}
An optimization problem was formulated using four typical days for the REC system to find the optimal number of solar PV and BESS for individual participants to minimize electricity bills. Formally, the problem was stated as
\begin{equation}
\begin{aligned}
    & \max_{\mathcal{X}}
    && \underbrace{\textrm{\eqref{forpoc2}}}_\text{Net Profit of REC}\\
    & \textrm{s.t.}
    && \underbrace{\textrm{\eqref{eqn:Loadbalance_REC2}}}_\text{Power balance constraints}\\
    &&& \underbrace{\textrm{\eqref{P number}, \eqref{eqn:ChargeDischarge_REC2} to \eqref{eqn:SOC3}, \eqref{B number}, \eqref{Positive}, \eqref{eqn:payback}}}_\text{Operational constraints}\footnotemark\\
\end{aligned}
\label{eqn:OP_2}
\end{equation}
\footnotetext{The state of charge periodicity was imposed as $S_{v\Tilde{T}} = S_{(v-1)\Tilde{T}}$, where $v \in \{1,2,3,4\}$ denoted the number of days and $\Tilde{T}$ denoted the number of samples in each day.}
where $\hat{\mathcal{X}} = \{N^{p,n}, N^{b,n}, P_{t}^{self,n}, P_{t}^{sell,n}, P_{t}^{sh,n}, P_{t}^{c,n}, P_{t}^{d,n}, \\ S_{t}^{n} ~|~ \forall t \in \hat{\mathcal{T}}, \forall n \in \mathcal{N}\}$ was the set of decision variables and $\hat{\mathcal{T}}:= \{1,2,\dots, \hat{T}\}$ was the set of hours over the four periodic seasonal days with a total number of hours $\hat{T}$.

\subsection{Individual Marginal Contribution}
This method calculated the marginal contribution of each participant based on their individual energy sharing profile, which quantified the unique contribution of each participant. For individual contribution, it was necessary to find the total incentive received from the public authority and the savings on self-consumption, which were expressed as
\begin{equation}
    I_{x}^{sh} = \sum_{t\in \hat{\mathcal{T}}}(C_{t}^{E} + C_{t}^{T})P_{t}^{sh}\Delta t.
\end{equation}
According to each individual participant's consumption, the distribution factor was expressed as
\begin{equation}
    \zeta^{n} = \frac{\sum_{t\in \hat{\mathcal{T}}}d_{t}^{n}\Delta t}{\sum_{n \in \mathcal{N}}\sum_{t\in \hat{\mathcal{T}}}d_{t}^{n}\Delta t},
    \label{eqn:Distribution_Factor}
\end{equation}
where $\zeta^{n}$ represented the proportion of the total incentive received by the $n^{\rm{th}}$ participant, proportional to their individual electricity consumption. The electricity bill of the $n^{\rm{th}}$ participant after joining the REC was then formulated as
\begin{equation}
    \overline{EB}^{n}  = EB^{n} - \beta (C_{x}^{sell,n} + R_{x}^{sav,n} + \zeta I_{x}^{sh,n}),
\end{equation}
where $EB$ was the annual electricity bill before joining the REC.

\section{Simulation Results}
The optimization algorithm was implemented using MATLAB software with the YALMIP toolbox \cite{YALMIP} and the Gurobi solver \cite{GUROBI}. The hardware used for the simulation was an Intel Core i7-11800H, 2.3GHz server with 32GB of RAM. The developed REC model was applied to a real case study to test its practical use and evaluate its potential. The case study was related to the low voltage grid in Roseto Valfortore, a small town and commune in the Foggia province in the Apulia region of southeastern Italy. For the simulation analysis, five household demand profiles of participants from the Roseto Valfortore energy community were collected, and historical solar irradiance and temperature data for the town of Roseto Valfortore were obtained from the online monitoring dashboard \cite{PVGIS}, with the sampling rate $\Delta t = 1$ h. Furthermore, the solar PV panel and BESS specifications were provided in Table \ref{tab: Specifications}, which were taken from \cite{SVT} and \cite{TESLA}, respectively. Furthermore, the electricity purchase rate $C_{t}^{P}$ for blocks F1, F2, and F3 was taken to be 0.195, 0.165 and 0.125 \euro / kWh, respectively. At the same time, the sale electricity rate $C_{t}^{S}$ for blocks F1, F2 and F3 was set to 0.075, 0.055 and 0.035 \euro / kWh, respectively. Furthermore, it was considered that the electricity sharing rate $C_{t}^{E}$ and the electricity transmission rate $C_{t}^{T}$ were 0.11 \euro / kWh and 0.00822 \euro / kWh, respectively. The simulation study assumed that: (1) the daily load profile of all participants was known; (2) the irradiance profile of the geographical location was known; and (3) the distribution network transmission line was considered ideal, i.e. no constraints were imposed on the network transmission line.
\begin{table}[h!]
\small
\centering
\caption{Specifications of the solar PV and BESS}
\label{tab: Specifications}
\setlength\tabcolsep{2.2pt}
\begin{tabular}{l|c|r}
\hline
Description                     & Abbr.          & Value      \\ 
\hline
\multicolumn{3}{c}{PV} \\
\hline
Rating of PV-panel              & $P^{r}$             & 0.43 kW         \\
Size of one panel               & $A^{p}$             & 2.4 m$^{2}$    \\
PV-panel lifespan               & $L^{p}$             & 25 years        \\
Power temperature coefficient   & $\gamma$            & 0.043 \%/$^\circ$C  \\
Cost of solar PV                & $\hat{C}^{p}$       & 1200 \euro/kW       \\
Irradiance during STC           & $\hat{E}$           & 1 kW/m$^{2}$     \\
Temperature during STC          & $\hat{\theta}$      & 25$^\circ$C      \\
Operating \& maintenance costs  & $\textrm{OPEX}^{p}$ & 24\euro/kWh/year  \\
TF contract agreement costs     & $\textrm{CA}^{p}$   & 1\euro/kWh/year
\\
\hline
\multicolumn{3}{c}{BESS} \\ 
\hline
Normalized Capacity             & $B$                 & 5 kWh   \\
Charging/Discharging efficiency & $\eta^{c},\eta^{d}$ & 90\%    \\
Min Charging/Discharging power  & $\underline{P}^{c}, \underline{P}^{d}$ & 0 kW    \\
Max Charging/Discharging power  & $\overline{P}^{c}, \overline{P}^{d}$   & 1.25 kW  \\
Min state of charge             & $\underline{S}$     & 0.5 kWh \\
Max state of charge             & $\overline{S}$      & 4.5 kWh \\
Cost to buy BESS                & $\hat{C}^{b}$       & 250\euro/kWh \\
BESS lifespan                   & $L^{b}$             & 12 years \\
Operating \& maintenance costs  & $\textrm{OPEX}^{b}$ & 24\euro/kWh/year  \\
TF contract agreement costs     & $\textrm{CA}^{b}$   & 1\euro/kWh/year  \\
\hline
\end{tabular}
\end{table}

\begin{figure*}[h!]
    \centering
    \includegraphics[height=5.3cm,width=17.4cm]{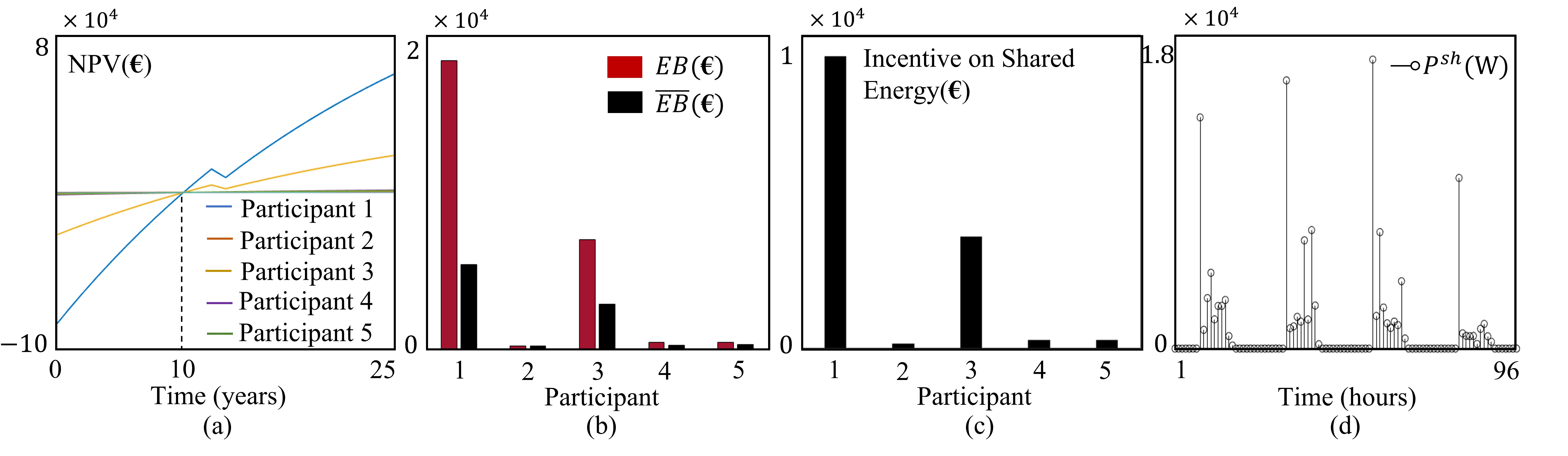}
    \caption{NPV, electricity bills, incentives of individual participants on shared energy and total amount of energy shared within the REC.}
    \label{fig:NPV_EB}
\end{figure*}
The optimal number of solar PV and BESS obtained for the five individual participants mentioned in Table \ref{tab:NpNb}.
\begin{table}[h!]
\centering
\caption{Optimal number of PV and BESS}
\label{tab:NpNb}
\setlength\tabcolsep{11.6pt}
\begin{tabular}{l|c|c|c|c|c}
\hline
Participants $\rightarrow$ & 1   & 2 & 3  & 4 & 5 \\ \hline
Solar PV     & 146 & 0 & 45 & 3 & 2  \\ \hline
BESS         & 16   & 0 & 6  & 0 & 0 \\ \hline
\end{tabular}
\end{table}

As shown in Fig. \ref{fig:NPV_EB}(a), the NPV of individual participants was recovered within the payback period, which depended on the duration for which the participants wanted to recover their investment cost. It also suggested that investing in the REC was financially advantageous in the long run. Another observation was that the first and third participants replaced their BESS after 12 years, and therefore the graph showed a slight decline followed by an increase again because of the revenue generation and incentives from shared energy within the REC. Fig. \ref{fig:NPV_EB}(b) compared the electricity bills of the participants before and after joining the REC and showed that all participants reduced their electricity costs by becoming part of the community. Fig. \ref{fig:NPV_EB}(c) showed that the incentive for each participant was based on the amount of energy they contributed to the shared energy, highlighting the direct rewards for active energy sharing. Finally, Fig. \ref{fig:NPV_EB}(d) presented the total shared energy profiles for four representative days (one for each season), illustrating how the share of energy varied and highlighting seasonal differences.
\begin{figure*}[h!]
    \centering
    \includegraphics[height=13.4cm,width=17.55cm]{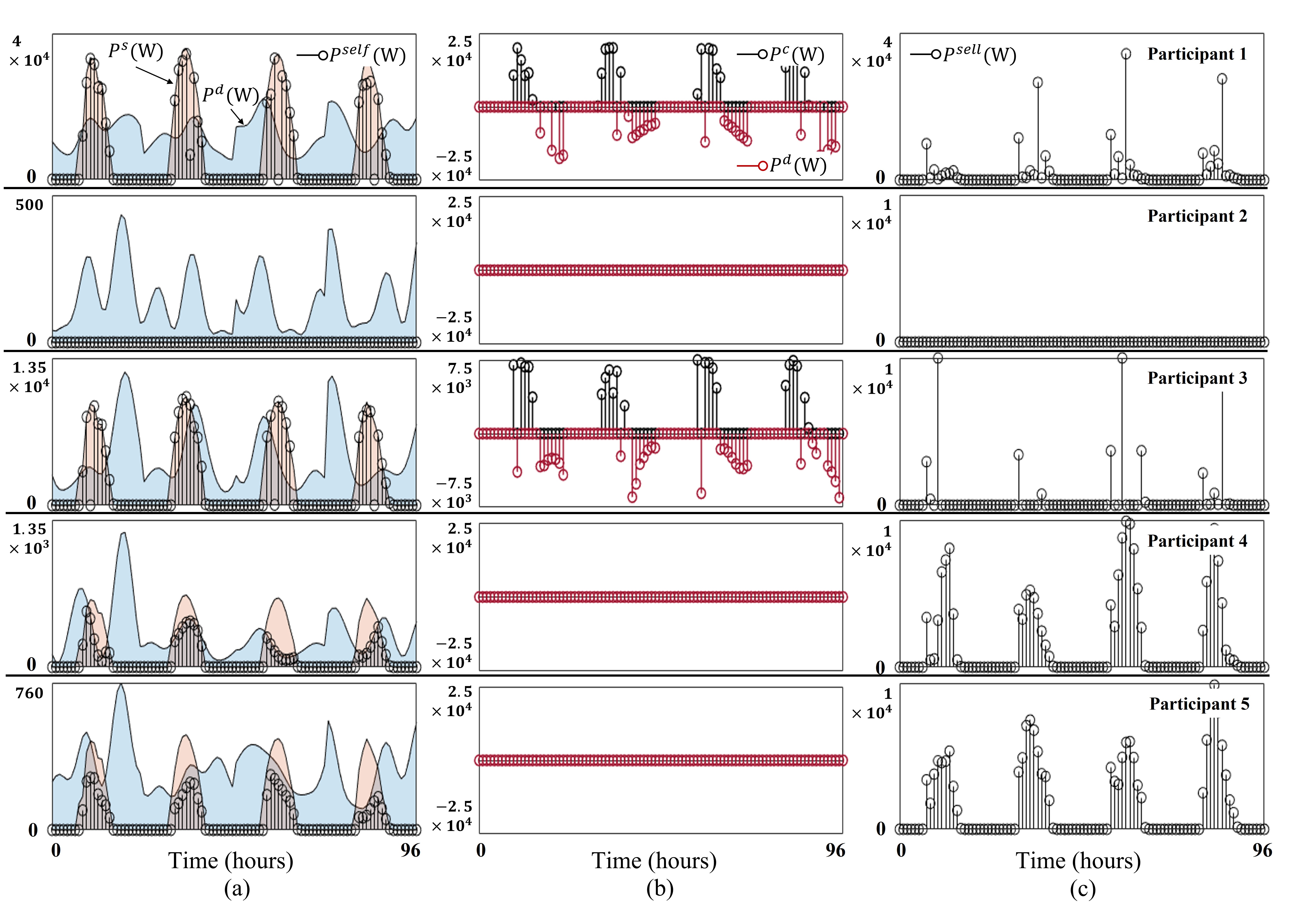}
    \caption{Four periodic days profile for demand, generation, self-consumption, charging/discharging power, and selling power for individual participants.}
    \label{fig:All_Figures}
\end{figure*}

Fig. \ref{fig:All_Figures} provided a detailed look at how five participants in the REC managed their electricity consumption, solar generation, and potential battery storage. Each row corresponded to one participant and the three columns (a), (b), and (c) highlighted key aspects of their energy flow. In Fig. \ref{fig:All_Figures}(a), the light blue area showed the participant’s electricity demand, while the light orange area represented their solar generation. The vertical stem line on top of the demand curve indicated the portion of electricity directly self-consumed from solar, illustrating how much of their demand was met by their generation in real-time. Focusing on each participant, it was observed that self-consumption ($P_{t}^{self}$) varied throughout the day as solar production ($P_{t}^{s}$) fluctuated. During peak sunlight hours, the stem line that indicated self-consumption ($P_{t}^{self}$) often matched a significant portion of the participant’s demand (light blue area), thereby reducing the dependence on the grid. Solar generation (light orange area) sometimes exceeded the demand curve, hinting at surplus electricity that could potentially be stored or sold to the grid. The degree of overlap between solar generation and demand was an essential indicator of how effectively participants used their renewable resources.

Fig. \ref{fig:All_Figures}(b) showed the charging behavior ($P_{t}^{c}$) and the discharging behavior ($P_{t}^{d}$) of the BESS. For participants who did not have a BESS installed, these plots remained at zero, reflecting no BESS activity. For those with a BESS, the positive values represented battery charging (using excess solar energy), while the negative values corresponded to discharge (supplying the participant's demand when solar production was insufficient or during periods of high grid cost). This battery operation allowed participants to optimize their solar use by storing energy for later consumption, thus increasing self-consumption and reducing peak grid demand. In Fig. \ref{fig:All_Figures}(c), the surplus power sold to the grid by each participant is depicted. When solar generation exceeded self-demand and battery charging needs, excess electricity was sold back to the grid. Participants without a BESS often showed more frequent or higher spikes in grid export during sunny periods because they had no means of storing surplus energy, while participants with batteries exported less at noon since they first charged their storage and then sold power to the grid once the battery was full or when it became economically optimal to discharge. These results showed that the REC participants used different strategies. Those with BESS and PV reduced their reliance on the grid and improved their financial returns. Even participants without BESS benefited by using solar power to reduce their dependence on the grid and by selling extra power back to the grid. In general, the results proved that the coordinated smart control strategy improved the economic and environmental performance of the REC.

\section{Conclusion}
This article proposed the optimal sizing formulation for individual REC participants using MINLP. Since simulation for a longer planning horizon was ineffective in solving, daily representative signals were used for each season of the year to determine the optimal number of solar PV and BESS with the objective of maximizing NP for individual REC participants. The FISTA algorithm was used to obtain these daily representative signals. Moreover, after receiving payment from the TF via the public authority, the fair revenue-sharing mechanism was proposed using the individual marginal contribution method. The simulation results showed that the participant received an additional incentive to participate in the REC. This incentive for individual participants was determined after making payment to the TF for real-time energy management, operation, and maintenance costs for installed solar PV and BESS. These findings highlighted the potential of a collaborative energy community to support the adoption of renewable energy while ensuring fairness for all participants. This work offered a clear and practical roadmap for creating an efficient, sustainable, and fair REC. In the future, this research could have been explored by applying this model in different settings and with other renewable energy sources.

\section*{Authorship contribution statement}
Vishal Kachhad: Conceptualization, Methodology, Software,
Validation, Formal Analysis, Investigation, and Data Curation. Amit Joshi: Methodology, Validation, Resources,
and Writing Reviews \& Editing. Luigi Glielmo: Supervision, Validation, Resources \& Writing reviews.
\section*{Declaration of competing interest}
The authors declare that they have no known competing
financial interests or personal relationships that could have
appeared to influence the work reported in this document.

\section*{Abbreviations}{
The following abbreviations are used in this manuscript:\\

\noindent 
\begin{tabular}{@{}ll}
PV    & Photovoltaics\\
BESS  & Battery Energy Storage Systems\\
FISTA & Fast Iterative Shrinkage-Thresholding Algorithm\\
NOCT  & Nominal Operating Cell Temperature\\
MINLP & Mixed-Integer Linear Programming\\
STC   & Standard Test Conditions \\
NPV   & Net Present Value \\
REC   & Renewable Energy Community \\
TF    & Technical Facilitator \\
RLA   & Rooftop Lease Agreement \\
PCC   & Point of Common Coupling \\
CF    & Cash Flows \\
OPEX  & Operating and Maintenance Costs
\end{tabular}
}

\end{document}